\documentclass[prl,twocolumn,showpacs,amsmath,amssymb]{revtex4-1}
\usepackage{graphicx} 
\usepackage{epsfig}
\usepackage{dcolumn} 
\usepackage{color}
\newcommand{\beq}{\begin{equation}}
\newcommand{\eeq}{\end{equation}}
\newcommand{\beqa}{\begin{eqnarray}}
\newcommand{\eeqa}{\end{eqnarray}}

\newcommand{\kvec}{{\bf k}}

\DeclareMathOperator{\sgn}{sgn}

\begin{document}
\title{Phase separation from electron confinement at oxide interfaces}
\author{N. Scopigno$^1$, D. Bucheli$^1$, S. Caprara$^{1,2}$, J. Biscaras$^3$, N. Bergeal$^3$, J. Lesueur$^3$,
and M. Grilli$^{1,2}$}
\affiliation{$^1$Dipartimento di Fisica, Universit\`a di 
Roma ``La Sapienza'', P.$^{le}$ Aldo Moro 5, 00185 Roma, Italy}
\affiliation{$^2$ISC-CNR and Consorzio Nazionale Interuniversitario per le Scienze Fisiche della 
Materia, Unit\`a di Roma ``Sapienza''}
\affiliation{$^3$Laboratoire de Physique et dÕEtude des Mat\'eriaux -
CNRS-ESPCI ParisTech-UPMC, PSL Research University, 10 Rue Vauquelin - 75005 Paris, France.}

\begin{abstract}
Oxide heterostructures are of great interest both for fundamental and applicative reasons. In particular
 the two-dimensional electron gas at the  LaAlO$_3$/SrTiO$_3$ or LaTiO$_3$/SrTiO$_3$ interfaces 
 displays many different physical properties 
and functionalities. However there are clear indications that the interface electronic state is
strongly inhomogeneous and therefore it is crucially relevant to investigate possible intrinsic 
electronic mechanisms underlying this inhomogeneity.
Here the electrostatic potential confining the electron gas at the interface is calculated self-consistently, 
finding that the electron confinement at
the interface may induce phase separation, to avoid a thermodynamically unstable state with a negative 
compressibility. This provides a generic robust and intrinsic mechanism for the experimentally observed inhomogeneous character of these 
interfaces. 
\end{abstract}
\date{\today}
\maketitle

The two-dimensional electron gas (2DEG) that forms at the interface of two insulating oxides, 
like LaAlO$_3$/SrTiO$_3$ and LaTiO$_3$/SrTiO$_3$ (hereafter generically referred to as LXO/STO)
\cite{ohtomo,Mannhart:2008uj,Mannhart:2010ha,Hwang:2012nm}, exhibits a rich phenomenology, such as a 
gate-tunable metal-to-superconductor transition \cite{reyren,triscone,espci1,espci2}, a magnetic-field-tuned 
quantum criticality \cite{espciNM}, and inhomogeneous magnetic responses \cite{ariando,luli,bert,metha1,metha2,bert2012}.
Tunneling \cite{richter,BCG} and SQUID magnetometry \cite{Kalisky} provide clear evidence of an inhomogeneous interface on 
both micro- and nanoscopic scales. Transport measurements report further signs of inhomogeneity and a percolative 
metal-to-superconductor transition with a sizable fraction of the 2DEG never becoming superconducting down to the lowest 
accessible temperatures \cite{CGBC,BCCG,caprara,SPIN}. For both fundamental reasons and applicative purposes, like device design, 
it is crucial to identify possible {\it intrinsic} mechanisms that may render the 2DEG so strongly inhomogeneous 
via a phase separation (PS). This is precisely the focus of the present work.

Here, we identify a very effective electron-driven mechanism leading to PS, based on the confinement of the 2DEG at the 
interface. From customary self-consistent calculations of the confining potential well in semiconductors, it is well 
known \cite{eisenstein} that a finite lateral extension usually renders the 2DEG more compressible than 
its strictly 2D counterpart. This effect is much stronger in LXO/STO than in ordinary semiconductor 
interfaces, due to the huge dielectric constant of STO, allowing for much larger electron densities, with a 
strong amplification of the self-consistent adjustments of the confining potential. As a consequence, a non-rigid 
band structure arises, which varies with the local electron density: an increased electron density is accompanied by a 
corresponding increase of the positive countercharges (due to oxygen vacancies and/or polarity catastrophe 
\cite{notacharges,nakagawa,chen,deluca,digennaro,zunger}),
from which the interfacial electrons are introduced and restoring the overall charge neutrality. For small-to-moderate increases 
of electron and countercharge densities the potential well deepens and the electron energy levels are shifted downwards. In 
this Letter we show that this mechanism leads {\it by itself} to PS. 

\begin{figure}[htbp]
\includegraphics[angle=0,scale=0.275]{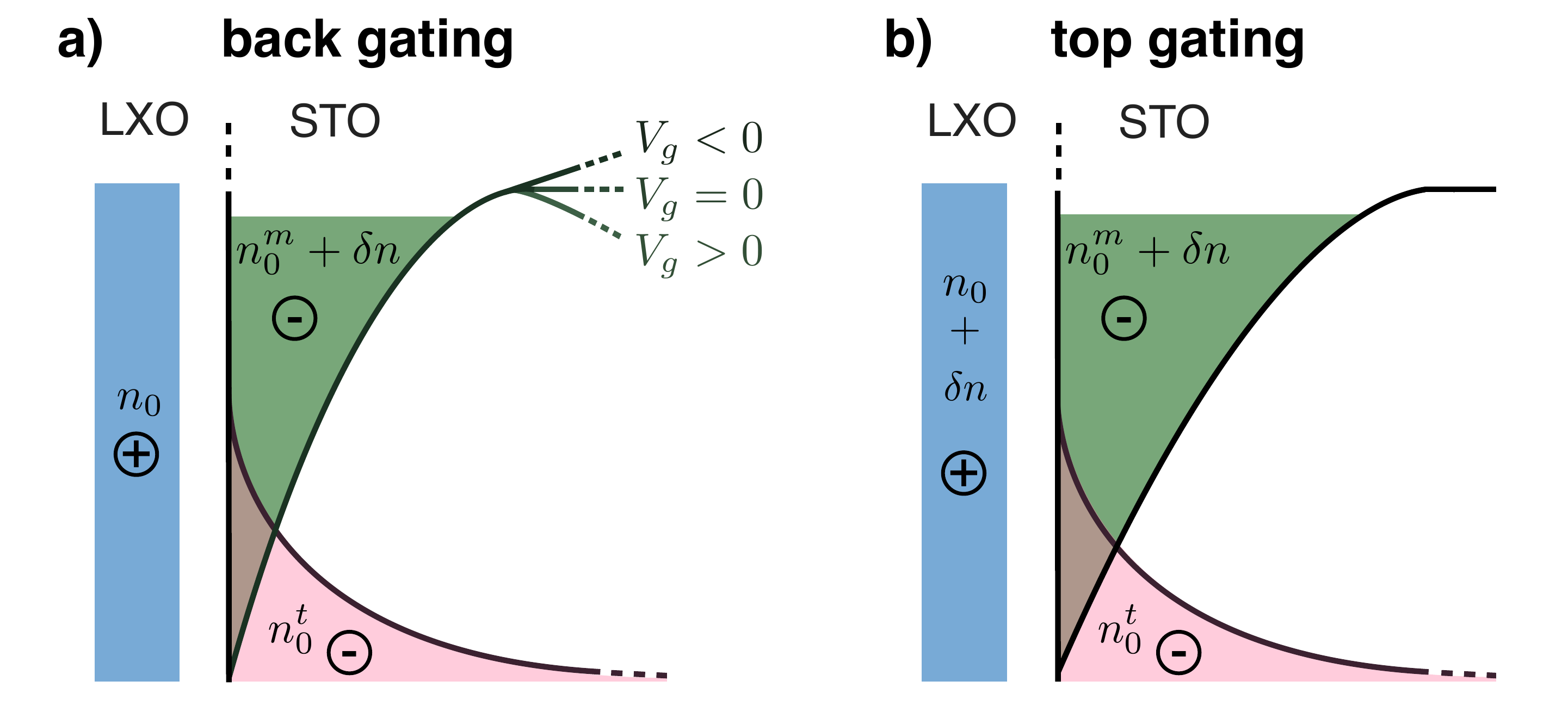}
\caption{Sketch of the interface for back (a) and top (b) gating. 
The confining potential depends on both mobile ($n_0^m$) and trapped ($n_0^t$) charges. 
Applying a positive (negative) voltage, $\delta n$ electrons per unit cell are added to (subtracted from) 
the interface and the potential changes accordingly.}
\label{schemes}
\end{figure}

{\it --- The model ---}
The thermodynamic stability of the system is investigated by varying the density of the interfacial gas {\it while
keeping the overall neutrality}. Therefore a corresponding amount of positive countercharges has to be varied (see Fig.\,\ref{schemes}).
Due to this tight connection between positive 
and negative charges the (in)stability will be determined by calculating the chemical potential of the whole system (i.e., of 
both the mobile electrons and of the rest of the charges). 
While for the electron part we will solve the quantum problem of the mobile electrons in the self-consistent 
confining well, the countercharges, the fraction of electrons trapped in impurity states of the bulk (see below) and the 
boundary conditions fixing the gating potential will determine the classical electrostatic energy of the system. We will 
keep all these contributions (for a detailed description see the Appendix) into account to calculate the total energy 
$\mathcal E$ and, in turn, the chemical potential $\mu= \mathcal E(N+1)-\mathcal E(N)\approx\partial_N\mathcal E$ (here $N$ 
represents the number of electrons, which is always kept equal to the number of countercharges).

The energy of the 2DEG is obtained through the calculation of the energy spectra as a function of 
the electron density, $n\equiv\delta n+n_0$ [henceforth, carrier densities, e.g., 
$n$, carrying no $z$ dependence are meant per interfacial unit cell (u.c.), and are related to their 
$z$-dependent counterparts by relations like, e.g., $n=\int_0^{\infty}dz\,n(z)$].
Here, the two contributions have different origin: $\delta n$ is the 
extrinsic component introduced by gating, while the intrinsic component $n_0$ \cite{notan0} originates 
from the electronic reconstruction due to the polarity catastrophe and/or from oxygen vacancies within the LXO 
layer. Which of the two dominates is not important in our calculations and we represent the related positive
countercharges as the light-blue shaded areas in the LXO side of Fig.\,\ref{schemes}.

What matters, instead, is the way the extrinsic charges are introduced, particularly in the case 
of back-gating [Fig.\,\ref{schemes}\,(a)]: Applying a positive voltage $V_g$, the electrostatic potential, 
after increasing in the region close to the interface, decreases linearly with distance, once the interfacial charge density 
has been exhausted [the electrons reside on the STO side, which we assume to occupy 
the $z>0$ half space and $n(z)\to 0$ for large $z$]. Then the electrons closest to the top of the well are weakly confined and some of them 
may escape and get trapped into the defects of bulk STO, as it is demonstrated by irreversibility effects
under large back-gating potentials \cite{baignoire}. Thus in the absence of trapped charges, 
the quantum well is intrinsically unstable upon positive back-gating. 
In the top-gating configuration [Fig.\,\ref{schemes}\,(b)], 
the leakage also occurs, because the Fermi energy of the electrons, attracted to the interface by the 
positive $V_g$, exceeds the confining potential on the STO side. In both configurations, we are therefore led to 
introduce trapped charges that we describe by a distribution $n_0^t(z)=(n_0^t/\lambda){\mathrm e}^{-z/\lambda}$ on the STO 
side, decaying over a distance $\lambda$ of several tens of nanometers (see pink shaded area in Fig.\,\ref{schemes}). This 
has the main effect of deepening the confining potential well, the Fermi energy being located substantially below its top 
(see, e.g., Refs.\,\onlinecite{espci2,meevasana}).

The mobile electrons occupy energy levels that are quantized in the $z$ direction and form a 2D 
band structure in the $xy$ (interfacial) plane. The electrostatic potential $\phi(z)$ confining the electrons is 
determined self-consistently with the mobile electron density distribution $n^m(z)\equiv n_0^m(z)+\delta n(z)$ 
(at external gating $V_g$), for a frozen distribution of trapped charges $n_0^t(z)$. The $z$ component of the 
factorized wave function $\Psi(x,y,z) = \zeta(z) \psi_{k_x k_y}(x,y)$ is the solution of the Schr\"odinger equation
yielding the sub-band energy $\varepsilon_i$,
\begin{equation}
\left[\frac{\hbar^2}{2m^z_\alpha}\frac{d^2}{dz^2} + e\phi(z)+\varepsilon_{i\alpha}\right]\zeta_{i\alpha}(z) = 0,
\label{Schrodinger:envelope}
\end{equation}
where the electron charge is $-e$ and $i=1,2,3,...$ is the sub-band index. The index $\alpha=xy,xz,yz$ labels the three Ti $t_{2g}$ 
orbitals, $d_{xy},d_{xz},d_{yz}$, where the electron 
mostly reside. The full 2D band structure is
$$
\varepsilon_{i,\alpha,\kvec} = \frac{\hbar^2 k_x^2}{2 m^x_\alpha} +\frac{\hbar^2 k_y^2}{2 m^y_\alpha} 
+ \Delta_\alpha+\varepsilon_{i\alpha}.
$$
Taking $\Delta_{xy}=0$, the energy offset $\Delta_{xz,yz}\equiv\Delta$ of the $d_{xz,(yz)}$ bands is experimentally found to be 
$\Delta\approx 50$\,meV \cite{salluzzo}. Similar values are found in first-principle calculations between the 
highest occupied $d_{xy}$ sub-band and the lowest $d_{xz,yz}$ sub-bands \cite{held}. Here we have to adjust 
$\Delta\approx 0-10$\,meV in order to self-consistently obtain the energy difference between the $d_{xy}$ and the 
$d_{xz,yz}$ sub-bands of order $50-60$\,meV (see the Appendix), showing that the energy offset mostly arises from the different 
mass along the $z$ direction.  We take the masses of the various bands 
as $m^{x,y}_{xy}=m_l$, $m^z_{xy}=m_h$;   $m^{x,z}_{xz}=m_l$, $m^y_{xz}=m_h$;  $m^{y,z}_{yz}=m_l$, $m^x_{yz}=m_h$.  
According to standard values [\onlinecite{santander}], we take $m_l=0.7\,m_e$ and $m_h=14\,m_e$ ($m_e$ is the electron mass).
Assuming full translational invariance along the $xy$ planes and integrating over a u.c. 
of area $a^2$ with a suitable normalization of $\psi_{k_x k_y}(x,y)$, the density of mobile electrons, 
at temperature $T=0$, reads 
\[
n^m(z)= 
\sum_{i\alpha} \lvert \zeta_{i\alpha}(z)\rvert^2
\int_{-\infty}^{\varepsilon_F}{d\varepsilon \, g_{i\alpha}(\varepsilon)},
\]
where $g_{i\alpha}(\varepsilon)=a^2(\pi \hbar^2)^{-1}\sqrt{m^x_\alpha m^y_\alpha}
\,\theta(\varepsilon-\varepsilon_{i\alpha}-\Delta_\alpha)$ is
the density of states (DOS) of the various $t_{2g}$ sub-bands, $\theta(\varepsilon)$ 
is the Heaviside function, and $\varepsilon_F$ is the Fermi energy. 

The electron distribution corresponds to an electrostatic potential $\tilde{\phi}(z)$ obeying the Poisson 
equation: 
\begin{equation}
\frac{\epsilon_0 a^2}{e}\frac{d}{dz}\left[\epsilon_r(E) \frac{d}{dz}\tilde{\phi}(z)\right]
= n^m(z)+n_0^t(z).\label{P_E}
\end{equation}
Here, the dielectric constant is a function of the electric field $E=-d\tilde{\phi}/dz$ via the
relation $\epsilon_r (E)= \left(A + B\lvert E \rvert \right)^{-1} + \epsilon_{\infty}$,
where \emph{A}, \emph{B}, and $\epsilon_{\infty}$ are experimentally measured constants \cite{Neville}. Owing to 
the nearly ferroelectric character of STO, $\epsilon_r$ can reach very large values ($\gtrsim 25\times 10^3$) but, 
due to the very strong interfacial electric field \cite{espci2,CPG}, we find that, near $z=0$, 
$\epsilon_r\approx\epsilon_{\infty}\approx 100-300$. The crucial point of the above derivation is that the calculation is 
self-consistent only if the two potentials, $\phi$ from Eq. (\ref{Schrodinger:envelope}) and $\tilde{\phi}$ from Eq. (\ref{P_E}), 
coincide. 

Besides the difficulties stemming from the self-consistency, there are additional subtleties coming from the 
boundary conditions, which vary for the back- or top-gating configuration. In the former case we fix
a density of positive charges $n_0=\int_0^{\infty}dz\,[n_0^m(z)+n_0^t(z)]$ (per u.c.) at $z=0^-$. 
We made this simplifying choice, which turns out to be slightly less favorable to the occurrence of PS,
to avoid the distinction between the case of oxygen vacancies (for which the countercharge would be uniformly distributed 
in the LXO) and the polarity catastrophe (in which the countercharges suitably distribute themselves in the polar planes 
of LXO in order to minimize the energy \cite{kopp}). 
Having fixed the positive charges, we consider the electric field at $z=0^+$ (i.e., the 
slope of the confining potential) at the interface. The electric field deep inside the STO 
[where $n_0(z)\rightarrow 0$] is fully determined by the gate potential because the intrinsic electronic density 
and the corresponding positive charges in the LXO compensate and have thus no effect in this region. On the other 
hand, the electrons coming from back-gating create a field $V_g/L$, with 
$e\delta n= \epsilon_0a^2\int_0^{V_g/L}\epsilon_r (E)dE$, where $L$ is the thickness of the STO substrate.
In the top-gating configuration, instead, the density of positive charges at $z=0^-$ amounts to $n_0+\delta n$, and 
the electric field vanishes deep inside the STO substrate.

Once the quantum problem of mobile electrons is solved, the rest of the energy is due to the electrostatic contribution 
(per u.c.) of electric fields due to the overall distribution of the mobile ($m$) 
and the fixed ($f$, from gates and trapped) charges
\beq
\mathcal E_{es}=\frac{\epsilon_0a^2}{2}\int_{-\infty}^\infty \epsilon_r(E)\left[ E^{f\,2}(z)-E^{m\,2}(z)\right]dz,
\label{electrostatic}
\eeq
with $E=E^f+E^m$.
Notice that, since the Hartree-like electrostatic energy is double-counted in the quantum Hamiltonian,
the contribution of the mobile charges must be subtracted in Eq. (\ref{electrostatic}).
In the Appendix we provide details of how the above fields are calculated.

\begin{figure}
\vspace{0 truecm}
\includegraphics[angle=0,scale=0.3]{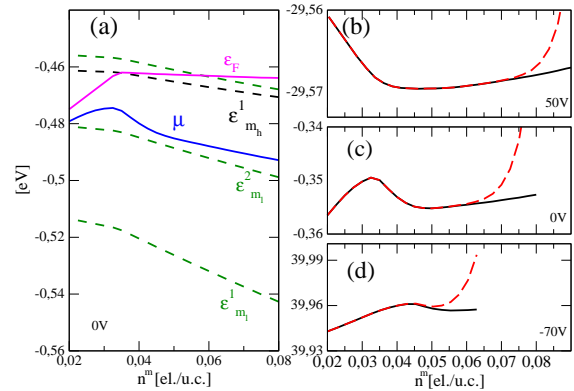}
\caption{(a) Fermi energy (magenta solid line), electron-only (see text) chemical potential (blue solid line),
sub-band levels (green dashed for the $d_{xy}$ and black dashed for the $d_{xz,yz}$ levels)
 as a function of the build-in density $n_0$ at zero gating potential. (b) Chemical potential 
 as a function of the mobile electron  density $n^m=n_0^m+\delta n$  at fixed values of the back gating potential 
$V_g \propto \delta n$ (the $\delta n$ electrons due to gating are thus also fixed). 
The black solid (red dashed) lines correspond to the solution in the absence (presence) of a countercharge background with
short-range rigidity. The short-range background contribution to the chemical potential is taken of the form 
$\mu_{sr}(n^m)=A\left[\left(n^m-\delta n \right)/0.065\right]^{p}$, where the exponent $p=19$ measures the short-range rigidity and $A=7\times 10^{-4}$.
For $V_G=50 \, V$, $\delta n=0.013668$.
(c) Same as (b) with $\delta n=0$. (d) Same as (b) with $\delta n=-0.01711$.}
\label{muvsn0}
\end{figure}

{\it --- Results---}
To evidentiate that the electron confinement is the driving mechanism of PS we first consider the pure quantum problem
neglecting the classical electrostatic contributions to the free energy. 
We thus report in Fig 2(a) the evolution of the different sub-bands levels as a function of the mobile electron density
$ n^m$, which is the sum of the initial as grown carrier density $n_0^m$ and the extra charges 
$\delta n$ added by electrostatic gating. This evolution is a direct consequence of the non-rigidity of the bands. 
The resulting Fermi energy $E_F$ is going up as expected, while the chemical potential $\mu^m$ of the mobile electrons 
has a non-monotonic behavior. Indeed, when $E_F$ crosses the first heavy band 
($d_{xz,yz}$ which have the largest DOS along the $x-y$ plane), $\mu^m$ starts to decrease. 
The most important result of this paper is that, for a given range of mobile electron densities, 
the chemical potential decreases while electrons are added, resulting into a negative compressibility.
 The entailed thermodynamic instability is prevented by PS, 
with the consequent inhomogeneous redistribution of electrons and countercharges. Remarkably,
the electron system tends to stay unstable (i.e. it has a negative compressibility) up to very large 
densities showing that this instability mechanism is strong and robust.
However, this cannot be a physical situation, since the compressibility remains negative, preventing a 
stable phase separation to occur. Indeed, this model is not complete. One has to introduce 
the energetic cost of the electrostatic fields generated by  all the charges in the 
system according to Eq.(\ref{electrostatic}). This leads to the chemical potential displayed in the panels 
of Fig. \ref{muvsn0} (b-d)  (black solid line) when $n^m$ is increased (the gate voltage is kept constant in this
 calculation, while the intrinsic $n_0^m$ is varied)..
Taking into account the quantum and electrostatic energies restores a physical situation, where the negative compressibility 
occurs in a limited range of doping, since for high $n^m$, $\mu$ increases when adding carriers.

\begin{figure}
\vspace{0 truecm}
\includegraphics[angle=0,scale=0.25]{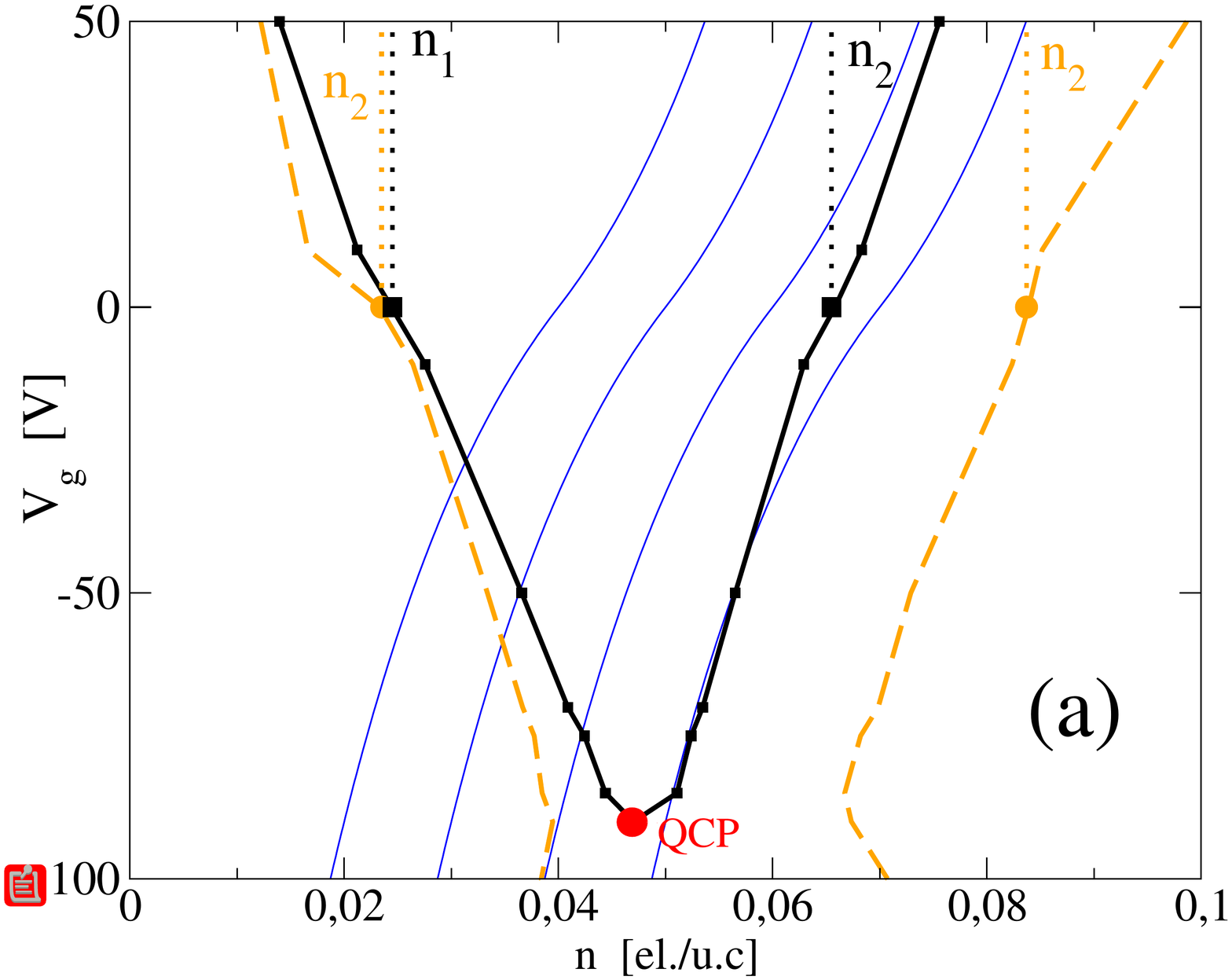}
\includegraphics[angle=0,scale=0.25]{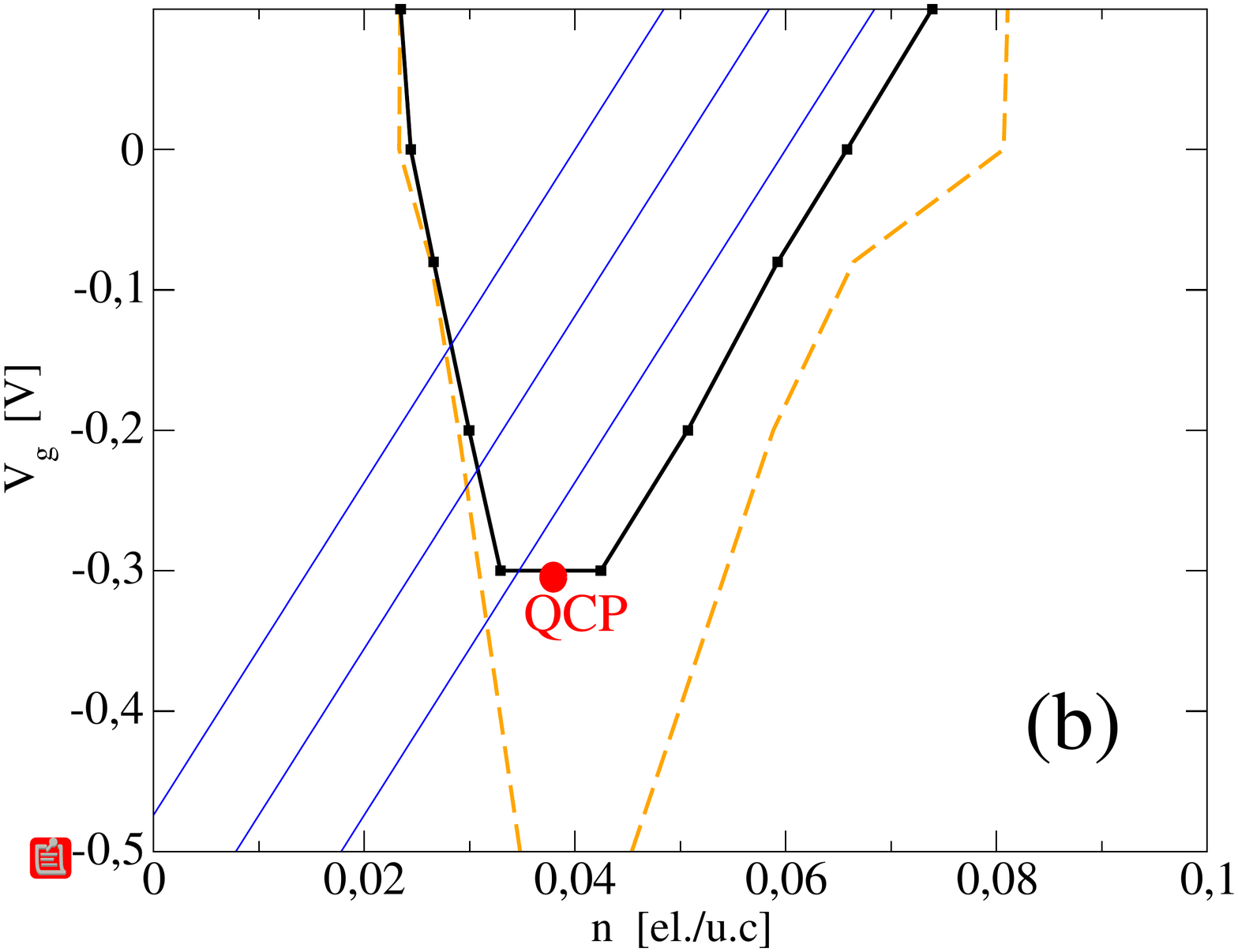}
\caption{(a) Gate potential versus mobile electron density phase diagram for the back gated LXO/STO interface in the absence (orange dashed lines) and in
the presence (black solid lines) of short-range rigidity of the countercharges. $n_1$ and $n_2$ are determined by
the Maxwell construction. The blue thin solid lines correspond to  $n_0=0.04,\,0.05,\,0.06,\,0.07$\,el/u.c. at $V_g=0$.
(b) Same as (a) for the top-gating case, with $n_0=0.04,\,0.05,\,0.06$\,el/u.c. 
}
\label{dome}
\end{figure}

Now the increasing classic electrostatic energy cost tend to stabilize the system at large $n_0^m$ and allow to 
perform a standard Maxwell construction on the $\mu$ vs. $n^m$ curves at each gate potential. 
The resulting phase diagram is shown in Fig \,\ref{dome}) for back [Fig \,\ref{dome}(a)] and top 
[Fig \,\ref{dome}(b)] gate.  The dashed orange lines limit the stability region (in between the system is phase separated), 
corresponding to the densities $n_1$ and $n_2$. We also plot the gate voltage as a function of $n_m$ 
(blue solid lines) for different samples having different densities at zero gating.

As can be seen in Fig. \,\ref{dome} the "miscibility gap" remains open at very negative gate voltage, especially in the 
back-voltage configuration. This does not correspond to the experimental situations \cite{Hurand2015}. This discrepancy 
arises because, while we took into account the long-range electrostatic 
cost of the charge distribution, we did not include in our model any short range repulsion between the countercharges
(oxygen vacancies or  positive charges left by the polarity catastrophe reconstruction). 
While some mild fluctuations of electrons and compensating countercharges are accepted, when too many electrons would segregate carrying 
along too large densities of countercharges, the system becomes very rigid and the charge segregation stops.

To overcome this problem, we added a short-range repulsion to the total energy.
Of course a precise estimate of the energy involved in these 
repulsive interactions is out of reach for the present calculations and we only implement a phenomenological energy barrier 
that induces a rapid grow in $\mu(n)$ when electron densities above those experimentally found (with a maximum of $0.5$\,el/u.c., 
including the trapped charges) is reached. Due to this additional short-range mechanism, the PS region determined by
the Maxwell construction on the dashed red lines of Figs. \ref{muvsn0} (b-d) is reduced (black solid lines and squares in Fig. \ref{dome}) 
and, quite remarkably, it ends with a critical point at some critical negative gate value, $V_g^c$. The resulting phase diagram 
for back-gating is reported in Fig.\,\ref{dome}\,(a). The initial (as-grown) density $n_0$ is determined by several 
specific parameters (like, e.g., the number of LXO planes). 
Starting from a given $n_0$, the total density per u.c., $n=n_0+\delta n$, is then changed by the 
gating following the thin blue trajectories in the phase diagram. Inspection of these trajectories reveals a very 
large range of intrinsic densities $n_0$ leading to PS. Upon increasing the negative gating the overall average density 
$n$ decreases and the fraction of the system with lower density $n_1$ increases, in agreement with transport measurements 
\cite{caprara,SPIN,SUST}, until the system exits the PS region at some negative voltage. A similar behavior is found for the 
top-gating case whose phase diagram is reported in Fig.\,\ref{dome}(b).

{\it --- Discussion ---}
From Figs.\,\ref{dome} (a) and (b) one can see that there exists a broad range of the intrinsic density $n_0$ yielding 
a negative compressibility, prevented by PS. This PS yields an inhomogeneous 2DEG with associated inhomogeneity of the 
oxygen vacancies and/or electronic reconstructions. We carried out a detailed analysis (see also the Appendix) 
to identify the specific mechanisms determining this instability finding that it arises from two main features peculiar 
to these oxide interfaces.

First of all, the electrons at the interface are confined in the STO side where a large dielectric constant 
$(\epsilon_r > \epsilon_{\infty}\gtrsim 100)$ efficiently screens the electric fields. This allows for 
the accommodation of a large number of electrons ($\approx 10^{13}$\,cm$^{-2}$) on several confined levels. The large 
DOS coming from the contributions of the individual sub-bands greatly enhances the electron compressibility and facilitates 
the instability. This effect is stronger when $d_{xz,yz}$ sub-bands start to be filled because they have a rather
large DOS. The relevance of the $d_{xz},d_{yz}$ levels has already been asserted in Hartree-Fock calculations \cite{millis} 
and seems to be experimentally supported \cite{joshua}. The filling of these levels typically starts when the 
system enters the PS dome upon increasing $V_g$, and rapidly leads to increasingly more abundant regions with locally 
higher electron density. We speculate that this corresponds to the observed increase of high-mobility carriers and onset 
of superconductivity \cite{espci2,caprara}.

Secondly, a larger $n_0$ must correspond to larger density of countercharges on the LXO side, 
which attracts the interface electrons and deepens the confining potential well, causing a downward shift of the 
quantized levels. 

Interestingly, in a rather large range of $n_0$ values, the system exits the PS dome
in the vicinity of the critical point located here at $n_0^c\approx 0.0475$\,el/u.c. and $V_g^c\approx -70$\,V for back-gating
(see blue line trajectories in Fig.\ref{dome}). This suggests that decreasing $V_g$ the electrons in the 
LXO/STO interface eventually display some signatures of critical behavior where superconductivity will likely be 
affected by the strong quantum density fluctuations \cite{QCP-SC}.

In conclusion, we identified a mechanism of electron-driven PS. While the details of the phase-separated
region also depend on the short-range rigidity of the system, the existence and robustness of the PS is on 
a firm ground and can be responsible for the strong inhomogeneity observed at LXO/STO interfaces. This 
mechanism can also cooperate with other intrinsic \cite{CPG} or extrinsic (e.g., defects or domain walls \cite{Kalisky,bristowe})
mechanisms.
~
\vspace{0.5cm}
\par\noindent
We thank V. Brosco, C. Castellani, R. Raimondi, and G. Seibold for stimulating discussions.
We acknowledge financial support from the Sapienza
University Project n. C26A115HTN, the CNRS PICS program ``S2S'', the ANR JCJC ``Nano-SO2DEG'', the 
SESAME, OXYMORE and CNano programs from R\'egion IdF.

\vspace{2 truecm}

\begin{appendix}

\centerline{\bf APPENDIX}

\section{Numerical calculation of the self-consistent electronic well}

\begin{figure}
 \centering
 \includegraphics[scale = 0.50]{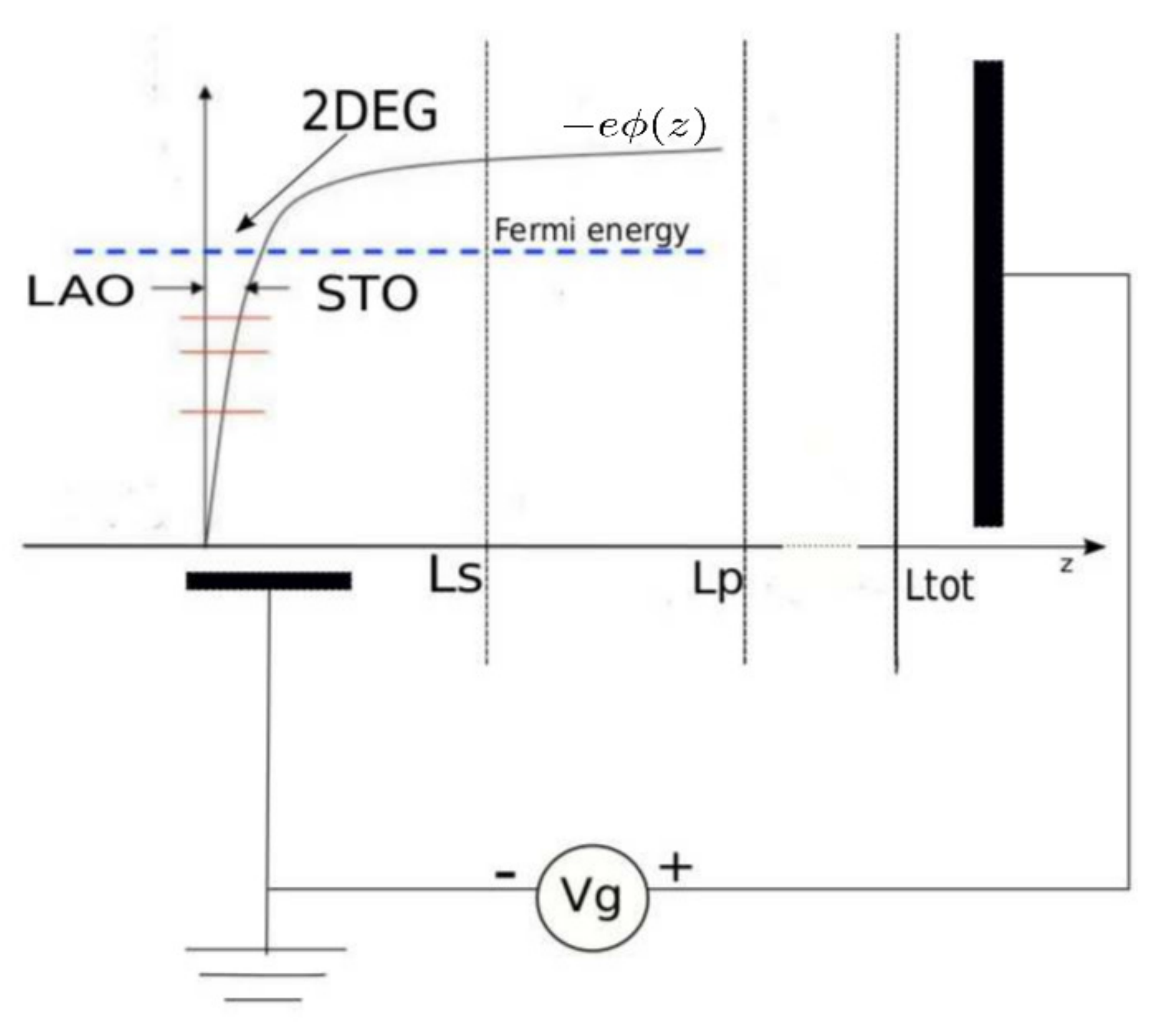}
 \caption{Schematic view of the back-gating configuration in LAO/STO.}
 \label{fig:STO_side_scheme}
\end{figure}

A well-tested approach to an eigenvalue problem in electronic quantum wells at the interface of two semiconductors is reported in \cite{stern_howard}. It consist 
in the simultaneous solution of the coupled Poisson and Schr\"odinger (in the effective mass approximation) equations 
\begin{eqnarray}
\left[\frac{\hbar^2}{2m_z}\frac{d^2}{dz^2} + e\phi(z)+\varepsilon_i\right]\zeta_i(z) &=& 0,\quad i=1,2,3, ...  \label{Schrodinger:envelope} \\
\frac{\epsilon_0 a^2}{e}\frac{d}{dz}\left[\epsilon_r(E) \frac{d}{dz}\phi(z)\right] &=& n(z) \label{P_E}
\end{eqnarray}
which gives us both the eigenfunctions and the confining potential well.
The close analogies between semiconductors and oxides heterointerfaces allows us to apply this method to the LXO/STO potential as well.
The boundary conditions  for the set (\ref{Schrodinger:envelope})-(\ref{P_E}) can be understood considering the schematic picture of the interface 
shown in fig. (\ref{fig:STO_side_scheme}).
Within this approach, the electric field and the displacement field are by symmetry on;y directed along the $z$ axis. Therefore henceforth we 
will indicate with $E$ and $D$ only the $z$ components of the fields.

The energy gap between the LXO and STO conduction bands is about $3eV$, so that in the low-filling regime it is safely to approximate the potential step at the 
LXO/STO interface as an infinite energy barrier. In addition the eigenfunctions have to be normalizable, therefore we are led to
set of boundary conditions
\begin{equation}
\zeta_i(0) = 0 \hspace{1cm} \zeta_i(\infty) = 0. \label{Schroedinger:BC_NoNum}
\end{equation}

The second condition can be implemented numerically exploiting the fact we are dealing with a confining potential: in a region far enough from the interface the 
eigenfunctions are exponentially decreasing. If $L_S$ resides in that region, the set (\ref{Schroedinger:BC_NoNum}) can be replaced by
\begin{equation}
\zeta_i(0) = 0 \hspace{1cm} \zeta_i(L_S) = 0. \label{Schroedinger:BC}
\end{equation}

The boundary conditions for the Poisson equation depend on both the (positive) countercharges left in the LXO valence band and on the gating voltage.
Experimental evidences show that the width of the 2DEG is only few nanometers, while the trapped charges extend over a 
few tens of nanometers \cite{espci2,baignoire}.
If $L_P$ is the thickness of the region in which all the charge is confinedwe can solve the Poisson equation analytically in the interval $(L_P,L_{TOT})$ and find appropriate boundary conditions by
the reasoning reported below. We take $L_P = 100 nm$, that is large enough to study also long-tail distributions
of the trapped charges.

Let us consider the interval $[0^{-},+\infty)$: in this region the system is neutral and we can reasonably impose the condition 

\begin{equation}
 E(-\infty) = E(+\infty) = 0 \label{E:neutrality}
\end{equation}
on the electric field, for any gating potential $V_G$.
In $ [0^+,+\infty)$ the neutrality does no longer hold, because the positive charges 
(left at the interface of LXO by the polar catastrophe or homogeneously distributed in the LXO layer if they are given by the oxygen vacancies) are now excluded. Thus
the electric field verifies the condition 

\begin{equation}
\int_{0^+}^{+\infty}{dz \frac{d}{dz}(\epsilon_r E)} = + \frac{n_0}{\epsilon_0} \Rightarrow \epsilon_r(0^+)E(0^+) = -\frac{n_0}{\epsilon_0}
\end{equation}
where $n_0$ are the polar catastrophe charges and we have used Eq. (\ref{E:neutrality}) to cancel the term $\epsilon_r(+\infty)E(+\infty)$.
Since all the electronic charges (mobile $n^m = n^m_0 + \delta n$, and trapped $n_0^t$) are between $0^+$ and $L_P$, the electric field in $z \in \left( L_P,L_{TOT} \right)$ is uniform even for the non linear experimental form of the relative dielectric constant $\epsilon_r = (A + B|E|)^{-1} + \epsilon_{\infty}$ [$A$, $B$ and $\epsilon_{\infty}$ are experimentally measured constants \cite{Neville}]. The boundary value for the potential thus reads
\begin{equation}
 \phi(L_P) = \frac{V_g}{L_{TOT}} L_P,
\end{equation}
where $V_g$ is the external gating.

\subsection{Numerical solution}
\label{numerical_solution}

Because of the nonlinear behavior of (\ref{Schrodinger:envelope})-(\ref{P_E}), an analytic approach is not viable and we adopted a numerical iterative procedure
to performs the calculation through the following steps:

\begin{itemize}
 \item The starting point is the potential at the \emph{nth} iteration, $\phi_n(z) = \phi_{old}^n(z)$, and his derivative, the electric field $E_n(z)=E_{old}^n(z)$,
 $z \in [0,L_S]$. We solve Eq. (\ref{Schrodinger:envelope}) using a Finite Element Method (\emph{FEM}) algorithm \cite{ram-mohan}, 
 finding the energies $\epsilon_i$ and the 
 eigenfunctions $\zeta_i$ which verify the conditions (\ref{Schroedinger:BC}). We recall that $L_S$ has to be
 taken far enough from the interface to ensure that the eigenfunctions are in a region in which they have an evanescent behavior. This can be done choosing different
 values for $L_S$ and checking them \emph{a posteriori}.
 The envelope functions are normalized imposing the condition $\int_0^{L_S}{\rvert \zeta(z) \lvert ^2 dz} = 1$.
 
 \item From the knowledge of the eigenvalues, the Fermi level can be calculated inverting numerically (e.g, using a bisection method) the relation 
 $n^m = \int_{-\epsilon_0}^{\epsilon_F}{g(\epsilon)d\epsilon}$, and from $n^m(z)= \sum_{i} \lvert \zeta_{i}(z)\rvert^2 \int_{-\infty}^{\varepsilon_F}{d\varepsilon \, 
 g_{i}(\varepsilon)}$ we have the density profile along \emph{z} of the 2DEG.
 Here only the mobile electrons contribute to define the Fermi energy since the trapped charges reside in impurity levels, deep or localized enough to be safely considered as a 
 stable quenched charge distribution.
 
 \item The profile of the potential well is found once one knows both the distribution of the 2DEG and the trapped (negative) charges. While the first is calculated
 self-consistently, the second is fixed at the zero\emph{th} step. There are no experimental
 evidences that determine univocally the distribution of the trapped charges, but a reasonable choice is a decreasing exponential distribution of the form
 \begin{equation}
  n_{0}^t(z) = \left(\frac{n_0^t}{\lambda}\right) e^{-\frac{z}{\lambda}}.
 \end{equation}
 Its shape is controlled by two parameters: the decay length $\lambda$ and the maximum value $n_0^t / \lambda$.
 
 \item The Poisson equation is solved in the interval $[0^+,L_P]$.  This gives a new potential $\phi(z) = \phi_{new}^n(z)$ 
 and a new electric field $E(z)=E_{new}^n(z)$.
 
 \item The update at the next step is performed defining the quantities
 \begin{eqnarray*}
  \phi_{n+1}(z) &=& s \phi_{old}^n(z) + (1- s)\phi_{new}^n(z) \\
  E_{n+1}(z) &=& s E_{old}^n(z) + (1-s) E_{new}^n(z)
 \end{eqnarray*}
  where the parameter $s \in [0,1]$ is used to control the difference between the old potential and the new one; 
  this reduces the risk that the iterative procedure escapes from the self-consistent solution and does not converge.
  The calculation stops as soon as the condition  
  \begin{equation}
   \int_0^{L_P}{\rvert \phi_{n+1} - \phi_{n} \lvert ^2 dz} < \delta \label{stop_condition}
  \end{equation}
  is fulfilled, for a suitably chosen accuracy $\delta$.
\end{itemize}

Through the previous five steps we obtain the potential well $V(z) = -e \phi(z)$ (with the corresponding Fermi energy $\epsilon_F$),
the electric field $E(z)$ and the density $n^m(z)$ of the 2DEG for a given value of $n_0$.

In principle the numerical solution may depend on the choice of the error $\delta$ in (\ref{stop_condition}), the initial potential $\phi_0(z)$
and the discretization length (according to the standard FEM \cite{ram-mohan}, we discretized in  $N_{elem}$  intervals 
the region in which we solve the Schr\"odinger equation). We performed different tests to ensure the reliability of the numerical solution.
In order to exclude the dependence on the initial potential, we varied the initial condition $\phi_0$ in a reasonable class of functions and kept 
$n_0$ and $V_g$ fixed. We find that the numerical solution is stable with respect to the initial data.
$N_{elem}$ and $\delta$ has been fixed to reach the uniform convergence. 
The solution is almost independent of $\delta$ while the choice of the first parameter is critical to obtain 
the uniform convergence.
In our work we took $N_{elem}$ so that $ \sup_{n_0} \left| \epsilon_F(N_{elem} +1) - \epsilon_F(N_{elem}) \right|$ is of order $\approx 10^{-5}$, 
a suitable request to reduce the error under the typical variations ($\sim 10^{-3} eV$) of the chemical potential $\mu(n)$.

\section{Electrostatic energy}
While the previous section solved the quantum problem of electrons in a self-consistent  potential, here we provide details of the calculation of the 
other electrostatic contributions to the total energy of the system arising form the fixed charges (those on the gating electrodes and the charges trapped
in the impurity states inside the STO) and from the mobile charges at the interface.

We consider two possible gating configurations. For the back-gating, the near region is the interval $z\in \left[0,L_p\right]$,
while the far region is the $z\in \left(L_p,L_{tot}\right]$ interval. On the other hand, for the top-gating, while the so-called near region is
the same, the ``far'' region corresponds to the LXO side of the heterostructure and corresponds to the $z \in \left[-d,0\right)$ interval.

In the near region we solve the Poisson equation
\begin{equation}
 \frac{d}{dz}D(z) = n^m(z)+n_0^t(z) \label{total_field}
\end{equation}
to find the electric displacement field $D(z)=D^m(z)+D^f(z)$ due to the mobile and fixed charges. Iin this region the only fixed charges are
those trapped in the STO side, while the charges on the electrodes and the countercharges on the LXO side only enter as boundary conditions. 
The electric field is
\begin{eqnarray}
\label{totalE}
 E &=& \frac{-\sgn (D)\left[1+ \epsilon_{\infty}A - \sgn(D) B D\right]}{2 B
 \epsilon_0 \epsilon_{\infty}} \\
& +& \frac{\sqrt{\left[1+\epsilon_{\infty}A -\sgn(D) BD\right]^2 +\sgn(D)4\epsilon_{\infty}ABD} }{2 B
 \epsilon_0 \epsilon_{\infty}} \nonumber
\end{eqnarray}
The total electric displacement field is obtained by numerically integrating Eq.(\ref{total_field}), while the field due to the 
 fixed charges is analytically found from 
\begin{equation}
  \frac{d}{dz}D^f(z) = n_0^t(z). \label{fixed_displacement}
\end{equation}
The result is \begin{equation}
 D^f(z) = -\frac{-en_0^t}{2 a^2} \left( 1- 2 e^{-\frac{z}{\lambda}}\right) + \frac{e n_0}{2 a^2} - \frac{e \delta n}{2 a^2}
\end{equation}

Once $D$ and $D^f$ are known, the electric displacement due to the mobile charges is obtained from $D^m = D - D^f$.
The electric fields of the fixed and mobile charges are
\begin{equation}
\label{Efm}
 E^{f,m} = \frac{D^{f,m}}{\epsilon_0}\frac{A + B \lvert E\rvert}{1 + \epsilon_{\infty} \left(A + B \lvert E\rvert \right)}.
\end{equation}
Notice that the non-linearity of $\epsilon_r$ entails the dependence of $D^{f,m}$ and $E^{fm}$ on the total electric field $E$ given by
Eq.(\ref{totalE})

In the far region we have different expressions for the top- and back-gating configurations.
For the latter the far region coincides with the bulk of the STO substrate, where 
\begin{eqnarray}
D^m(z) = \frac{en_0^t}{2 a^2}  - \frac{e \delta n}{2 a^2} - \frac{e n_0}{2 a^2}\\
D^f(z) = -\frac{en_0^t}{2 a^2} + \frac{e n_0}{2 a^2} - \frac{e \delta n}{2 a^2}
\end{eqnarray}
are constant. The corresponding electric fields are then obtained by solving Eqs.(\ref{totalE}) and (\ref{Efm}).

The far region in the top-gating configuration is instead given by the whole LXO side, where the dielectric constant $\epsilon_{LXO}$
is really constant 
\begin{eqnarray}
 E^m(z) = \frac{e \delta n}{2 \epsilon_0 \epsilon_{LXO} a^2} -\frac{en_0^t}{2 \epsilon_0 \epsilon_{LXO} a^2}   + \frac{e n_0}{2 \epsilon_0 \epsilon_{LXO} a^2} \\
 E^f(z) = \frac{e \delta n}{2 \epsilon_0 \epsilon_{LXO} a^2} +\frac{en_0^t}{2 \epsilon_0 \epsilon_{LXO} a^2}   - \frac{e n_0}{2 \epsilon_0 \epsilon_{LXO} a^2}
\end{eqnarray}


\end{appendix}


\begin{thebibliography}{99}
\bibitem{notacharges}The origin of the interfacial electrons is still controversial. Besides the electronic reconstruction 
mechanism related to the polarity catastrophe \cite{nakagawa}, the occurrence of oxygen vacancies is widely considered  
and controversial\cite{chen,deluca,digennaro}. See also \cite{zunger}  and 
references therein.
\bibitem{notan0} 
The gate potential $V_g$ (and the associated density $\delta n$) are external parameters determining the 
thermodynamic state of the system. In contrast, the density $n_0$ and its fluctuations are determined 
by the internal stability of the system. Thus, establishing whether for a given value of $n_0$, say $n_0^{exp}$ 
(fixed by the chemistry of the sample), the 2DEG is stable or not, amounts to determine whether the chemical 
potential is an increasing or a decreasing function of $n_0$ in the vicinity of $n_0^{exp}$, while keeping 
$V_g\propto \delta n$ fixed.
\end{thebibliography}
\end{document}